\documentstyle[aps,preprint]{revtex}
\draft
\tighten
\preprint{\vbox{\hbox{RUB-TPII-18/96} \hbox{hep-ph/9709221}}}
\title{Semileptonic decay constants of octet baryons \\
in the chiral quark-soliton model\footnote{Dedicated to Professor
Richard Lemmer on the occasion of his 65th birthday.}}
\author{Hyun-Chul Kim,
Maxim V. Polyakov~\footnote{On leave of absence from PNPI, Gatchina,
St.Petersburg 188350, Russia}, 
Micha{\l} Prasza{\l}owicz
\footnote{On leave of absence from Institute of Physics,
Jagellonian University, Cracow, Poland}, 
and Klaus Goeke}
\address{ Institute for Theoretical Physics II,\\
Ruhr-University Bochum, \\ D-44780 Bochum, Germany}
\date{August, 1997}
\begin{document}
\maketitle
\begin{abstract}
Based on the recent study of the magnetic moments
and axial constants within the framework of
the chiral quark-soliton model, we investigate the baryon semileptonic
decay constants $(f_1,f_2)$ and $(g_1, g_2)$.  
Employing the relations between the diagonal transition matrix elements
and off-diagonal ones in the vector and axial-vector channels,
we obtain the ratios of baryon semileptonic decay constants 
$f_2/f_1$ and $g_1/f_1$.  The $F/D$ ratio is also discussed 
and found that the value predicted by 
the present model naturally lies between that of the Skyrme model and
that of the nonrelativistic quark model.  The singlet axial constant
$g^{(0)}_A$ can be expressed in terms of the $F/D$ ratio and $g^{(3)}_A$
in the present model and turns out to be small.  The results are compared 
with available experimental data and found to be in good 
agreement with them.    
In addition, the induced pseudotensor coupling constants
$g_2/f_1$ are calculated, the SU(3) symmetry breaking being considered.
The results indicate that the effect of SU(3) symmetry breaking
might play an important role for some decay modes in
hyperon semileptonic decay.
\end{abstract}
\section{Introduction}
Baryon semileptonic decays have played an important role in various
facets to understand the structure of baryons.
  For example,  they provide information on the
Cabibbo-Kobayashi-Maskawa (CKM) angles $|V_{ud}|$ and $|V_{us}|$
as well as the $F/D$ ratio.  
Recently, baryon semileptonic decays have gained new interest 
in respect of a series of experiments measuring 
the first moment of the spin-dependent structure function 
$g_1(x)$~\cite{Ashmanetal,SMC,E142,E143}, since it is related to 
SU(3) invariant matrix elements of hyperon semileptonic processes 
$F$ and $D$.  

A recent high-precision measurement of 
$\Sigma^{-}\rightarrow n+e^-+\bar{\nu}$~\cite{Hsuehetal} shows 
a hint that the effect of SU(3) symmetry breaking might be
important to describe baryon semileptonic decays. 
Another experiment measuring $g_1/f_1$ in 
$\Lambda\rightarrow p + e^- +\bar{\nu}$ was conducted by
J. Dworkin {\em et al.} with high 
statistics~\cite{Dworkinetal}.  The result of Ref.~\cite{Dworkinetal}
prefers the hypothesis that the weak magnetism is less than
the CVC prediction ($f_2/f_1=0.97$).  In fact, Ref.~\cite{Dworkinetal} 
obtained $f_2/f_1=0.15\pm 0.30$ at which the fit yields the
minimum of $\chi^2$.  This value is really far from the CVC hypothesis.  
Also from the theoretical point of view
there were already serious doubts about the strong postulate 
of exact SU(3) symmetry in the Cabibbo theory~\cite{DoHo,GaKie,DHK}.

The effects of SU(3) symmetry breaking in baryon semileptonic decays
have been extensively studied from various points of view
\cite{DHK,Roos,Avenarius,ES,Ratcliffe}.
Donoghue {\em et al.}~\cite{DHK} made a careful analysis of 
hyperon semileptonic decays, 
considering the pattern of symmetry breaking based on the 
quark model.  It was asserted in Ref.~\cite{DHK}
that SU(3) breaking comes from two sources: 
the mismatch of the wave functions for the quarks 
and the recoil effect.  However, Roos reanalyzed hyperon semileptonic
decays~\cite{Roos} including recent data and showed that the scheme of 
symmetry breaking by Donoghue {\em et al.} fails to fit correctly
the $g_1/f_1$ ratio for $\Lambda\rightarrow p + e^- +\bar{\nu}$.  
The mismatch of strange-quark wave functions worsens the fit.   
Only by introducing a substantial second-class axial coupling
$g_2$ in the $\Lambda\rightarrow p + e^- +\bar{\nu}$,  one
could fit the data~\cite{Roos}.    
Avenarius~\cite{Avenarius} studied also SU(3) symmetry breaking 
in semileptonic hyperon decays, based on the Ansatz that 
SU(3) symmetry in the polarization at the current quark level 
is kept while SU(3) symmetry at the constituent quark level is broken.  
 The result was that with SU(3) symmetry broken the 
$F/D$ ratio $(0.73\pm 0.09)$ 
turned out to be larger than that of the case of SU(3) symmetry
$(0.59\pm 0.02)$.      
 Ehrnsperger and Sch\"afer~\cite{ES} came to a rather
different conclusion.  They showed that the effects 
of SU(3) symmetry breaking lead to a reduction of the $F/D$ ratio
$(0.49\pm 0.08)$.      
 Quite recently, Ratcliffe~\cite{Ratcliffe} 
reexamined SU(3) symmetry breaking effects in hyperon semileptonic decays.  
What he obtained is that $F/D=0.582$ for an SU(3) symmetric fit and
$F/D=0.570$ for an SU(3) breaking fit.  The result of 
Ref.~\cite{Ratcliffe} indicates that the effects of SU(3) 
symmetry breaking is rather tiny.  
There have been also similar discussions related to the validity of
SU(3) symmetry in the context of the spin structure of the proton.
In particular, Lipkin strongly criticized the use of SU(3) symmetry
in studying the spin structure of the proton~\cite{Lipkin}.
The topic of SU(3) symmetry breaking in hyperon semileptonic 
decays seems to be evidently far from the settlement yet 
and very difficult to be analyzed without relying 
on particular Ans\"atze.  

Recently, we investigated the magnetic moments of the baryon octet
\cite{KimPoBloGo} and the axial constants of the nucleon~\cite{BloPraGo}
within the framework of the chiral quark-soliton model 
($\chi$QSM)~\cite{DPP}, taking into account the 
$1/N_c$ rotational corrections and linear 
$m_s$ corrections which furnish the effect of SU(3) symmetry breaking.
The magnetic moments and axial constants which are given in terms of
diagonal matrix elements can be, however, related to the off-diagonal 
transition form factors, {\em i.e.} semileptonic weak form factors 
$f_1$, $f_2$ and $g_1$.  Therefore, we can easily evaluate them, using former
calculations.  The presence of the $m_s$ corrections allows
the nonvanishing values of induced pseudotensor coupling constants
$g_2$.  

The large $N_c$ limit ($N_c\rightarrow\infty$) provides a useful guideline
in understanding the low-energy properties of the baryon 
systematically~\cite{Witten}, though in reality $N_c$ is equal to 3.  
In the large $N_c$, the nucleon can be viewed as a classical 
soliton of the pion field.  
 An example of the dynamical realization 
of this idea is  given by the Skyrme model~\cite{ANW}. 
However, the $\chi$QSM presents a more realistic picture than the 
Skyrme model.  In the light of chiral perturbation theory,
the effective chiral action on which the $\chi$QSM is based contains 
automatically the four-derivative Gasser--Leutwyler terms~\cite{chpt}
and the venerable Wess-Zumino term~\cite{WZ} 
with correct coefficients~\cite{DE,AF,DSW}.  Moreover, 
the $\chi$QSM interpolates between the Skyrme model and the
nonrelativistic model (NRQM)~\cite{PraBloGo,tensor}, because the
$\chi$QSM is ideologically close to the Skyrme model in the limit of 
large soliton size while as the size of the soliton approaches zero 
the $\chi$QSM reproduces the results of the NRQM.  

The aim of the present paper is to investigate the semileptonic 
decay constants $(f_1,\ f_2)$ and $(g_1,\ g_2)$ within the 
framework of the $\chi$QSM.

The outline of the paper is as follows: In the next section, 
we sketch briefly the basic formalism of the $\chi$QSM.     
In section III, we first discuss the semileptonic decay constants in
SU(3) flavor symmetry.  As it should be, it is shown that the
induced pseudotensor coupling constant $g_2(0)$ vanishes in
SU(3) symmetry within the framework of the $\chi$QSM.  
The $F/D$ ratio is also discussed.  The singlet axial constant
$g^{(0)}_A$ is shown to be related to the $F/D$ ratio and
the axial constant $g^{(3)}_{A}$.  
 Considering the strange quark mass $m_s$,
we discuss the effect of SU(3) symmetry breaking on the coupling
constants $f_1$, $f_2$ and $g_1$.  The $g_2$ is also evaluated.
We compare the results with those obtained in the case of
SU(3) symmetry.  The deviations from the Cabibbo theory are
discussed in detail.  In section IV, we summarize the present
work and draw the conclusion.

\section{General Formalism}
The transition matrix element ${\cal M}_{B_1\rightarrow B_2 l \bar{\nu}_l}$
for the process $B_1\rightarrow B_2 l \bar{\nu}_l$ can be written as
\begin{equation}
{\cal M}_{B_1\rightarrow B_2 l \bar{\nu}_l}\;=\; \frac{G}{\sqrt{2}}
\left(\begin{array}{c} V_{ud} \;(\mbox{for}\; \Delta S = 0)\\
V_{us} \;(\mbox{for}\; |\Delta S| = 1) \end{array}\right)
\langle B_2| J^W_{\mu}|B_1\rangle
\bar{u}_l (p_l) \gamma^\mu \left(1-\gamma_5\right)u_{\bar{\nu}_l}(p_\nu),
\end{equation}
where $G$ denotes the effective Fermi coupling constant and $V_{ud},\;
V_{us}$ stand for the Cabibbo-Kobayashi-Maskawa angles.  
The leptonic current $\bar{u}_l (p_l) \gamma^\mu \left
(1-\gamma_5\right)u_{\bar{\nu}_l}(p_\nu)$ is the known part.  
The hadronic weak current $J^W_{\mu}$ 
has following spin and flavor structures:
\begin{equation}
J^W_{\mu}\;=\;\left\{\begin{array}{c} 
\bar{\psi}(x) \gamma_\mu (1-\gamma_5) \frac12 (\lambda_1 \pm i \lambda_2)
\psi (x), \;\mbox{for}\; \Delta S = 0 \\
\bar{\psi}(x) \gamma_\mu (1-\gamma_5) \frac12 (\lambda_4 \pm i \lambda_5)
\psi (x), \;\mbox{for}\; |\Delta S| = 1.
\end{array}\right.
\end{equation}

The transition matrix element of the hadronic weak current
$\langle B_2| J^W_{\mu}|B_1\rangle$ can be expressed in terms of six 
independent form factors:
\begin{eqnarray}
\langle B_2| J^W_{\mu}|B_1\rangle&=&\bar{u}_{B_2} (p_2) \left[
\left\{f_1 (q^2) \gamma_\mu - \frac{i f_2 (q^2)}{M_1}
\sigma_{\mu\nu} q^\nu + \frac{f_3 (q^2)}{M_1} q_\mu\right\} \right.
\nonumber\\
&&-\left.\left\{g_1 (q^2) \gamma_\mu - \frac{i g_2 (q^2)}{M_1}
\sigma_{\mu\nu} q^\nu + \frac{g_3 (q^2)}{M_1} q_\mu\right\}\gamma_5
\right] u_{B_1} (p_1)
\label{Eq:formfac}
\end{eqnarray}
with the momentum transfer $q=p_2 - p_1$.
The form factors $f_i$ and $g_i$ are real quantities depending only
on the square of the momentum transfer in the case of 
$CP$-invariant processes.
 We can safely neglect $f_3$ and $g_3$ for the reason that 
on account of $q_\mu$ their contribution to the decay rate
is proportional to the ratio $\frac{m^2_{l}}{M^2_{1}}\ll 1$, 
where $m_l$ represents the mass of the lepton ($e$ or $\mu$) in the
final state and $M_1$ that of the baryon in the initial state .  

It is already well known how to deal with the hadronic matrix elements
such as $\langle B_2| J^W_{\mu}|B_1\rangle$ in the $\chi$QSM
(for a review see \cite{review}).  Hence, we shall briefly explain how to
calculate them with regard to the semileptonic processes .  
 The $\chi$QSM is characterized by a low-momenta QCD partition function
\cite{dp} which is given by a functional integral over $8$ pseudoscalar
and quark fields:
\begin{equation}
{\cal Z}\;=\;\int {\cal D}\psi {\cal D}\psi^\dagger {\cal D}\pi^a 
\exp \left(\int d^4 x \psi^\dagger 
\beta (- i \rlap{/}{\partial} + \hat{m} + MU^{\gamma_5}) \psi\right).
\label{Eq:part}
\end{equation}
Integrating out quark fields leave us with the effective chiral action
\begin{equation}
{\cal Z}\;=\; \int {\cal D}\pi^a \exp\left(-S_{\rm eff}[\pi] \right),
\end{equation}
where 
\begin{equation}
S_{\rm eff} \;=\; -{\rm Sp} \ln 
\beta (- i \rlap{/}{\partial} + \hat{m} + MU^{\gamma_5})
\end{equation}
with the pseudoscalar chiral field
\begin{equation}
U^{\gamma_5}\;=\; \exp{(i\pi^a \lambda^a \gamma_5)} \;=\;
\frac{1+\gamma_5}{2}U\;+\;\frac{1-\gamma_5}{2}U^\dagger.
\end{equation}
$\hat{m}$ is the matrix of the current quark mass given by
\begin{equation}
\hat{m} \;=\;\mbox{diag} (m_u,m_d,m_s) 
\;=\; m_0{\bf 1} \;+\; m_8 \lambda_8.
\label{Eq:mass}
\end{equation}
$\lambda^a$ represent the usual Gell-Mann matrices normalized as 
$\mbox{tr}(\lambda^a\lambda^b) = 2 \delta^{ab}$.   Here, we have assumed
isospin symmetry ($m_u=m_d=\overline{m}$).  $M$ stands for the 
dynamical quark mass arising from the spontaneous chiral
symmetry breaking, which is in general momentum-dependent~\cite{dp}. 
We regard $M$ as a constant and employ the proper-time regularization
for convenience.  The $m_0$ and $m_8$ in Eq.~(\ref{Eq:mass}) 
are respectively defined by
\begin{equation}
m_0\;=\; \frac{2\overline{m}+m_s}{3}\;\simeq\; \frac{m_s}{3},
\;\;\;\;m_8\;=\; \frac{\overline{m}-m_s}{\sqrt{3}}\;
\simeq \; \frac{-m_s}{\sqrt{3}}.
\end{equation}
 The operator $i D$ is expressed in Euclidean space in terms of
the Euclidean time derivative $\partial_\tau$
and the Dirac one--particle Hamiltonian $H(U^{\gamma_5})$
\begin{equation}
i D \; = \; \partial_\tau \; + \; H(U^{\gamma_5}) 
+ \beta \hat{m} 
\label{Eq:Dirac} 
\end{equation}
with
\begin{equation}
H(U^{\gamma_5}) \; = \; \frac{\vec{\alpha}\cdot \nabla}{i} 
\;+\; \beta MU^{\gamma_5} 
\label{Eq:hamil}
\end{equation}
$\beta$ and $\vec{\alpha}$ are the well--known 
Dirac Hermitian matrices.  The $U$ is assumed to have a structure 
corresponding to the embedding of the SU(2)-hedgehog into SU(3):
\begin{equation}
U_c\;=\;\left(\begin{array}{cc}
U_0 & 0 \\ 0 & 1 \end{array} \right)
\end{equation}
with 
\begin{equation}
U_0\;=\; \exp \left(i \hat{r}\cdot \vec{\tau} P(r)\right).
\end{equation}
$P(r)$ is called profile function.
 The partition function of Eq.(\ref{Eq:part}) can be simplified
by the saddle point approximation which is exact in the large $N_c$ limit.
One ends up with a stationary profile function $P(r)$ which is 
evaluated by solving the Euler--Lagrange equation corresponding to
$\delta S_{\rm eff}/\delta P(r)=0$.  This gives a
static classical field $U_0$.  The soliton is quantized by introducing
collective coordinates corresponding to ${\rm SU(3)}_{\rm f}$ rotations 
of the soliton in flavor space (and simultaneously 
${\rm SU(2)}_{\rm spin}$ in spin space)
\begin{equation}
U (t, \vec{x}) \;=\; R(t) U_c (\vec{x}) R^\dagger (t),
\end{equation}
where $R(t)$ is a time-dependent SU(3) matrix.  The quantum states
from this quantization are identified with the SU(3) baryons 
according to their quantum numbers.  
In the large $N_c$ limit, the angular velocity of the soliton $\Omega
= R^\dagger (t) \dot{R} (t)$ can be regarded as a small parameter,
so that we can use it as an expansion parameter. 
After the rotation, the Dirac differential operator Eq.(\ref{Eq:Dirac})
can be expressed as
\begin{equation}
i\tilde{D}\;=\; \left[\partial_\tau + H(U^{\gamma_5})+ \Omega(t)
+\gamma_4 R^\dagger (t) \hat{m} R (t)\right].
\end{equation}
Then the propagator $(i\tilde{D})^{-1}$ can be expanded with regard to
the angular velocity $\Omega$ and the strange quark mass $m_s$:
\begin{equation}
\frac{1}{i\tilde{D}}\;\simeq\; \frac{1}{\omega +iH} - \frac{1}{\omega +iH}
\Omega \frac{1}{\omega +iH} - \frac{1}{\omega +iH}
\gamma_4 R^\dagger(t) \hat{m}R(t) \frac{1}{\omega +iH}.
\label{Eq:expand}
\end{equation}
The transition matrix element of the hadronic weak current given by 
Eq.(\ref{Eq:formfac}) can be related to the correlation function
\begin{equation}
\left\langle 0 \left| J_{B_1} (\vec{x}, \frac{T}{2}) 
\bar{\psi}\hat{\Gamma}{\hat{O}}
\psi J^{\dagger}_{B_2} (\vec{y}, -\frac{T}{2})\right|0\right\rangle
\end{equation}
at large Euclidean time T.  $\hat{\Gamma}$ and $\hat{O}$ are abbreviations
for the corresponding spin and flavor operators.
The $J_B$ denotes the baryon current which is constructed from $N_c$
quark fields
\begin{equation}
J_B=\frac{1}{N_c!}\varepsilon^{i_1 \ldots i_{N_c}} \Gamma^{\alpha_1
\ldots \alpha_{N_c}}_{SS_3II_3Y} \psi_{\alpha_1 i_1}\ldots
\psi_{\alpha_{N_c} i_{N_c}}.
\end{equation}
$\alpha_1 \ldots \alpha_{N_c}$ are spin--isospin indices, 
$i_1 \ldots i_{N_c}$ are color indices,
and the matrices $\Gamma^{\alpha_1 \ldots \alpha_{N_c}}_{SS_3II_3Y}$
are taken to endow the corresponding current with the quantum numbers
 $SS_3II_3Y$.  The $J_B(J^{\dagger}_B)$ plays a role of annihilating
(creating) the baryon state at large $T$.   
The rotational $1/N_c$ and linear $m_s$ corrections being taken into 
account as shown in Eq.(\ref{Eq:expand}), 
the relevant transition matrix elements can be written as follows:
\begin{eqnarray}
\left(f_1 + f_2\right)^{(B_1 \rightarrow B_2)} & = &
 w_1 \langle B_2 | D^{(8)}_{X3} | B_1 \rangle 
\;+ \;w_2 d_{pq3}
 \langle B_2 | D^{(8)}_{Xp}\, \hat{S}_q | B_1 \rangle 
\;+\; \frac{w_3}{\sqrt{3}}
 \langle B_2 | D^{(8)}_{X8}\, \hat{S}_3 | B_1 \rangle 
\nonumber \\
& + & m_s \left[\frac{w_4}{\sqrt{3}}
d_{pq3}\langle B_2 |D^{(8)}_{Xp}\, D^{(8)}_{8q}|B_1 \rangle 
+w_5 \langle B_2|\left(D^{(8)}_{X 3}\, D^{(8)}_{88} 
+ D^{(8)}_{X8}\, D^{(8)}_{83} \right)| B_1 \rangle \right. \nonumber\\
&+& w_6 \left. \langle B_2 |\left(D^{(8)}_{X 3}\, D^{(8)}_{8 8}
- D^{(8)}_{X 8}\, D^{(8)}_{83} \right)| B_1 \rangle 
\right]
\label{Eq:f1f2}
\end{eqnarray}
for the transition magnetic moments 
$(f_1(0) + f_2 (0))^{(B_1 \rightarrow B_2)}$ and
\begin{eqnarray}
g_1^{(B_1 \rightarrow B_2)} & = &
 a_1 \langle B_2 | D^{(8)}_{X3} | B_1 \rangle 
\;+ \;a_2 d_{pq3}
 \langle B_2 | D^{(8)}_{Xp}\, \hat{S}_q | B_1 \rangle 
\;+\; \frac{a_3}{\sqrt{3}}
 \langle B_2 | D^{(8)}_{X8}\, \hat{S}_3 | B_1 \rangle 
\nonumber \\
& + & m_s \left[\frac{a_4}{\sqrt{3}}
d_{pq3}\langle B_2 |D^{(8)}_{Xp}\, D^{(8)}_{8q}|B_1 \rangle 
+a_5 \langle B_2|\left(D^{(8)}_{X 3}\, D^{(8)}_{88} 
+ D^{(8)}_{X8}\, D^{(8)}_{83} \right)| B_1 \rangle \right. \nonumber\\
&+& a_6 \left. \langle B_2 |\left(D^{(8)}_{X 3}\, D^{(8)}_{8 8}
- D^{(8)}_{X 8}\, D^{(8)}_{83} \right)| B_1 \rangle 
\right],
\label{Eq:g1}
\end{eqnarray}
for the transition axial constants $g^{(B_1 \rightarrow B_2)}_1 (0)$.
The induced pseudotensor coupling constants 
$g_2^{(B_1 \rightarrow B_2)}$ are expressed by
\begin{equation}
\frac{g_2^{(B_1 \rightarrow B_2)}}{M_{B_1}}= 4 m_s
(\beta_1 i f_{ab3}+\beta_2 i \varepsilon_{ab3}) \cdot
 \langle B_2| D_{X a} D_{8 b}|B_1\rangle .
\label{g2}
\end{equation}
The parameters $w_i$, $a_i$ and $\beta_j$ depend on dynamics of the 
chiral soliton.  
As for the expressions for $w_i$ and $a_i$, one can find them in 
Ref.~\cite{KimBloPoGo} and Ref.~\cite{BloPraGo}, respectively.  
$\beta_1$ and $\beta_2$ can be written explicitly as~\cite{Pol90,weak}
\begin{eqnarray}
\beta_1+\beta_2 & = &
\frac{iN_c}{24\sqrt{3}} \int \frac{d \omega}{2 \pi}\, {\rm Sp}\left(
\frac{1}{\omega + i H }\gamma_4 \tau^i \frac{1}{\omega + i H }
i\varepsilon_{ijk }\tau^j x^k \gamma_5 \right),  \\
\beta_1&=&-\frac{iN_c}{12\sqrt{3}} \int \frac{d \omega}{2 \pi}\, 
{\rm Sp} \left(\left\{\frac{1}{\omega + i H }\gamma_4 , 
\frac{1}{\omega + i H_0 }\right\}
 \gamma_5 \vec{x} \cdot \vec{\tau}\right).
\end{eqnarray}

The remarkable feature of the soliton picture of the baryons
is that the singlet axial charge of the nucleon $g_A^{(0)}$ is expressed
in terms of the {\em same} parameters $a_i$ in Eq.(\ref{Eq:g1}):
\begin{equation}
g_A^{(0)} \;=\; \frac12 a_3 + \sqrt{3} m_s \left(a_5 - a_6\right).
\end{equation}
Hence, in this picture the value of $g_A^{(0)}$ can be extracted
by fitting the data on semileptonic decays {\em without} 
resorting to those on polarized deep inelastic scattering (see the next
section for the analysis of the SU(3)--symmetric case).    
With SU(3) symmetry explicitly 
broken by $m_s$, the collective Hamiltonian is no more
SU(3)--symmetric.  The pure octet states are mixed with the higher 
representations such as anti-decuplet states.
 Therefore, the baryon wave function with spin $S=1/2$ 
requires the modification due to the strange 
quark mass ($m_s$).  
 Since we treat $m_s$ perturbatively up to the first order,
the collective wave function of the baryon octet can be written as
\begin{equation}
\Psi_B (R) \;=\; \Psi^{(8)}_B (R) + m_{\rm s}\,c_{\overline{%
10}}^B\ \Psi^{(\overline{10})}_B (R)  
+m_{\rm s}\,c_{27}^B\ \Psi^{(27)}_B (R) ,
\label{Eq:waveftn}
\end{equation}   
where
\begin{equation}
c^{B}_{\overline{10}} \;=\; \frac{\sqrt{5}}{15}(\sigma - r_1)
\left[ \begin{array}{c}  1 \\ 0 \\ 1 \\ 0 
\end{array} \right] I_2 ,\;\;\;\;
c^{B}_{27} \;=\; \frac{1}{75}(3\sigma + r_1 - 4r_2)
\left[ \begin{array}{c} \sqrt{6} \\ 3 \\ 2 \\  \sqrt{6}
\end{array}\right] I_2 
\label{Eq:g2}
\end{equation}
in the basis of $[N,\ \Lambda,\ \Sigma,\ \Xi]$.
Here, $B$ denotes the SU(3) octet baryons with the spin 1/2.
The constant $\sigma$ is related to the nucleon 
sigma term $\Sigma\;=\;\overline{m}\langle N|
\bar{u}u+\bar{d}d |N\rangle\;=\;3/2 \overline{m} \sigma$ 
and $r_i$ designates
$K_i/I_i$, where $K_i$ stands for the anomalous moments of inertia
defined in~Ref.\ \cite{Blotzetal}.
The collective wave function can be explicitly written in terms of
the SU(3) Wigner $D^{({\mu})}$ function:
\begin{equation}
\Psi^{(\mu)}_B \;=\;  (-)^{S_3-1/2}\sqrt{\mbox{dim}(\mu)}
\left[D^{(\mu)}_{(YTT_3)(-1SS_3)} \right]^*.
\end{equation}

Hence, we have two different contributions of SU(3) symmetry breaking:
One from the effective Lagrangian and the other from the wave function
corrections.  All contributions of SU(3) symmetry breaking
are kept in linear order of $m_s$.  Apart from these two contributions,
we shall see in the next section that in the case of $f_2(0)$ 
the mass differences between octet baryons come into play.  
Hence, on the whole, we have three   
different sources for the SU(3) symmetry breaking in the present model.

\section{Semileptonic decay constants in the $\chi$QSM}
\subsection{Exact SU(3) symmetry}
In exact SU(3) symmetry,
the vector coupling constants $f_1(0)$ and $f_2(0)$ can be simply
expressed in terms of the anomalous magnetic moments of the 
proton and neutron. 
Similarly, $g_1(0)$ in all decay modes can be parametrized
in terms of two SU(3)-invariant constants $F$ and $D$~\cite{GaiSau}:
\begin{equation}
g^{(B_1\rightarrow B_2)}_1 = F C^{B_1\rightarrow B_2}_F
+ D C^{B_1\rightarrow B_2}_D,
\end{equation}
where $C^{B_1\rightarrow B_2}_F$ and
$ C_D^{B_1 \rightarrow B_2} $ are SU(3) Clebsch-Gordan coefficients
that appear when an octet operator is sandwiched between octet states.
The superscript refers to the hadrons involved and subscripts $F$ and $D$
denote the antisymmetric and symmetric parts.
We list the expressions for $f_1(0)$, $f_2(0)$ and $g_1(0)$ 
in Table I~\footnote{Note that the signs in the $\Lambda\rightarrow p$, 
and $\Sigma^-\rightarrow n$ modes are different from Ref.~\cite{GaiSau}.  
However, the ratios $f_2(0)/f_1 (0)$ and $g_1(0)/f_1(0)$ are not
affected.  We have employed the phase convention \` a la 
De Swart~\cite{Swart,Alfaro}.}.
  
The pseudotensor coupling constants $g_2(0)$ are all 
predicted to be zero in exact SU(3) symmetry 
because of $G$-parity.  In fact, it can be shown that 
$g_2(0)$ vanish in the present model.  
To do so, it is of great use to introduce a transformation
\begin{equation}
G_{5} = \tau_2 C\gamma_5,
\end{equation}
where $C$ is the operator of charge conjugation:
$C^{-1}\gamma_\mu C = -\gamma^{T}_\mu$.
Under this transformation, the one-body Dirac Hamiltonian,
Dirac and Pauli matrices are respectively changed as follows:
\begin{eqnarray}
G_5^{-1} H G_5 &=& H^T, \nonumber \\
G_5^{-1} \gamma_\alpha G_5 &=& \gamma^{T}_\alpha, 
\;\;\mbox{for} \; \alpha=1,\cdots,5 \nonumber \\
G_5^{-1} \tau_a  G_5 &=& -\tau^{T}_a .
\end{eqnarray}
The pertinent trace for the 
leading order contribution to $g_2(0)$ can be written as
\begin{equation}
{\rm tr} \left[\left\langle \vec{x} \left| \frac{1}{\omega + i H} \gamma_5 
\tau_a x_i \right| \vec{x} \right\rangle \right] .
\end{equation}
Utilizing the $G_5$ transformation and the properties of the trace
${\rm tr}(M^T)={\rm tr}(M)$ and ${\rm tr}(WMW^{-1})={\rm tr}(M)$, 
we can show that the trace of the leading contribution vanishes:
\begin{eqnarray}
{\rm tr} \left[\left\langle \vec{x}\left|\frac{1}{\omega + i H} \gamma_5 
\tau_a x_i \right| \vec{x} \right\rangle \right]&=&
{\rm tr} \left[\left\langle \vec{x}\left|G_5^{-1} 
\frac{1}{\omega + i H} \gamma_5 
\tau_a x_i G_5 \right| \vec{x} \right\rangle \right]^T \nonumber \\
&=&-{\rm tr} \left[\left\langle \vec{x}\left|
\frac{1}{\omega + i H} \gamma_5 \tau_a x_i \right| \vec{x} \right
\rangle \right].
\end{eqnarray}
Similarly we can prove that the $1/N_c$ rotational corrections to $g_2(0)$ 
also disappear.

In exact SU(3) symmetry, $g^{(B_1\rightarrow B_2)}_1$
can be written in terms of three independent dynamic quantities $a_i$
calculable in the present model:
\begin{equation}
g_1^{(B_1 \rightarrow B_2)} \; = \;
 a_1 \langle B_2 | D^{(8)}_{X3} | B_1 \rangle 
\;+ \;a_2 d_{pq3}
 \langle B_2 | D^{(8)}_{Xp}\, \hat{S}_q | B_1 \rangle 
\;+\; \frac{a_3}{\sqrt{3}} \langle B_2 | D^{(8)}_{X8}\, 
\hat{S}_3 | B_1 \rangle .
\label{Eq:g1sym}
\end{equation}
As for the process $n\rightarrow pe^-\bar{\nu}$, 
$g^{(n\rightarrow p)}_1$ becomes just the axial coupling constant
$g^{(3)}_{A}$.  
After some straightforward manipulation~\cite{BloPraGo},
we end up with the expression for $g^{(3)}_{A}$~\cite{WaYo,BloPoGo}
\begin{equation}
g^{(3)}_{A} \;=\; \frac{7}{30} \left(-a_1 + \frac12 a_2 + \frac{1}{14} 
a_3 \right).
\label{Eq:g3a}
\end{equation}
We can also obtain the singlet axial coupling constant $g^{(0)}_A$
\begin{equation}
g^{(0)}_A \;=\; \frac12 a_3.
\label{Eq:g0a}
\end{equation}
It is of great interest to see how the $\chi$QSM plays an interpolating
role between the Skyrme model and the NRQM.
In the limit of the NRQM the dynamic quantities 
$a_i$ in Eq.(\ref{Eq:g1sym}) are respectively~\cite{antidec,petrov}
\begin{equation}
a_1 = -5, \;\;\; a_2 = 4, \;\;\; a_3 = 2,
\label{Eq:anum}
\end{equation}
which give the correct values of $g^{(3)}_A$ and $g^{(0)}_{A}$
in the NRQM, {\em i.e.} $g^{(3)}_A=5/3$ and $g^{(0)}_{A}=1$.
The ratio $F/D$ then can be written in terms of $a_i$
\begin{equation}
\frac{F}{D} \;=\; \frac59 \left(\frac{-a_1 + \frac12 a_2 + \frac12 a_3}
{-a_1 + \frac12 a_2 - \frac16 a_3}\right).
\label{Eq:fd}
\end{equation}   
In the limit of the Skyrme model, {\em i.e.} when the size of the soliton
is very large~\cite{PraBloGo}, $g^{(0)}_{A}$ approaches zero or $a_3$
vanishes.  Hence, the $F/D$ ratio becomes obviously $5/9$ in this limit.
This result is exactly the same as what was obtained by Bijnens 
{\em et al.}~\cite{BSW} and Chemtob~\cite{Chemtob}.
On the other hand, in the limit of the NRQM, {\em i.e.}
in the limit of zero soliton size,
the present model gives the value of $F/D = 2/3$ which is exactly
the same value as that of the SU(6) NRQM.  
The present model predicts $F/D$ to be 0.61 -- correspondingly,
we have $g^{(0)}_A=0.36$~\cite{BloPoGo} -- which lies between 
the value from the Skyrme model $(5/9)$ and that from the NRQM $(2/3)$.
The $\chi$QSM shows here again interpolation between 
the Skyrme model and the NRQM~\cite{PraBloGo,tensor}.   
Notably, the smallness of the singlet axial charge $g^{(0)}_A$ 
is directly related to the fact that $F/D$ 
does not deviate much from that of the Skyrme model $(5/9)$,
where $g^{(0)}_A$ is known to be zero~\cite{BrodskyEllisKarliner}. 
Using Eqs.(\ref{Eq:g3a},\ref{Eq:g0a},\ref{Eq:fd}), we can express 
the singlet axial constant $g_A^{(0)}$ in terms of the $F/D$ ratio
and $g_A^{(3)}$:
\begin{equation}
g^{(0)}_A\;=\; \frac{9g^{(3)}_A}{1+F/D}\left(\frac{F}{D}-\frac59\right).
\end{equation}
Substituting the value of $F/D= 0.582$ obtained in a recent 
analysis~\cite{Ratcliffe} and $g^{(3)}_A = 1.26$, 
one gets $g^{(0)}_A=0.19$.

The $g_1$ is normally determined in experiments with $g_2$ assumed to
be zero.  However, Hsueh {\em et al.}~\cite{Hsuehetal}
extracted for the first time the induced pseudotensor coupling 
constant $g_2$ in $\Sigma^{-} \rightarrow n e^- \bar{\nu}$ decay.
This new experimental results give a reduced value for $g_1(0)$: 
Instead of $g_1(0)=0.328 \pm 0.019$ and $g_2=0$,
$g_1(0)=0.20 \pm 0.08$ and $ g_2(0)=-0.56 \pm 0.37 $ are obtained.
These results are remarkable, since they indicate that
the effect of SU(3) symmetry breaking might play an important role
in baryon semileptonic decays.  In fact, these new experimental
results have triggered discussions about the effect of SU(3) symmetry
breaking in hyperon semileptonic decays~\cite{Avenarius,Ratcliffe}.
Hence, we need to consider the explicit SU(3) symmetry breaking.

\subsection{SU(3) symmetry breaking effects}
As we have shown in section II, the strange quark mass $m_s$ 
provides the effect of SU(3) symmetry breaking in two 
different forms: One from the effective action and the other
from the wave function corrections.  Apart from these two contributions, 
the mass difference between baryon states must be considered 
in the case of $f_2(0)$.
   
By switching on SU(3) symmetry breaking to the first order in $m_s$, 
we obtain the expressions for $f_2(0)$ and $g_1(0)$ deviating from 
those listed in Table I.  It should be noted that by the Ademollo-Gatto
theorem~\cite{AdGa}
$f_1(0)$ do not get any contribution from linear $m_s$ corrections.
 We choose the combination of the magnetic moments 
in which all corrections from $1/N_c$ 
are cancelled to avoid ambiguity arising from the relations
between the hyperon magnetic moments~\cite{KimPoBloGo}. 
Hence we get the unambiguous relations between
hyperon magnetic moments and off--diagonal matrix elements:
\begin{eqnarray}
f_2^{(n \rightarrow p)}(0) &=& \frac12\left(\kappa_p-\kappa_n\right),
\nonumber\\
f_2^{(\Sigma^- \rightarrow \Sigma^0)}(0) &=& \frac{M_{\Sigma}}{2M_N}
\frac{1}{\sqrt2}\left(\kappa_{\Sigma^+}-\kappa_{\Sigma^-}\right), 
\nonumber\\
f_2^{(\Sigma^\pm \rightarrow \Lambda)}(0)&=& \mp
\frac{M_{\Sigma}}{\sqrt2 M_N} \mu_{\Sigma^0 \Lambda},\nonumber\\
\mu_{\Sigma^0 \Lambda}&=&\frac{1}{\sqrt3}\left(-\kappa_n+
\frac14 (\kappa_{\Sigma^+}+\kappa_{\Sigma^-}) -\kappa_{\Xi^0}+
\frac32\kappa_{\Lambda}\right), \nonumber\\
f_2^{(\Lambda \rightarrow p)}(0)&=& \frac{M_{\Lambda}}{2 \sqrt6 M_N}
\left(\kappa_n+\frac52 \kappa_p
+\frac12 \kappa_{\Sigma^-}-3\kappa_{\Lambda}-\kappa_{\Xi^-}\right),
\nonumber\\
f_2^{(\Sigma^- \rightarrow n)}(0)&=& \frac{M_{\Sigma}}{2 M_N}
\left(\kappa_n+\frac12 \kappa_p - \kappa_{\Sigma^-}
-\frac12 \kappa_{\Sigma^+}\right), \nonumber \\
f_2^{(\Xi^- \rightarrow \Lambda)}(0) &=& \frac{M_{\Xi}}{2 \sqrt6 M_N}
\left( \kappa_p -\frac12 \kappa_{\Sigma^+}+3\kappa_{\Lambda}-
\kappa_{\Xi^0}-\frac52 \kappa_{\Xi^-}\right),\nonumber\\
f_2^{(\Xi^- \rightarrow \Sigma^0)}(0)&=& \frac{M_{\Xi}}{2 \sqrt2 M_N}
\left( \kappa_{\Sigma^+}
+\frac12 \kappa_{\Sigma^-}-
\kappa_{\Xi^0}-\frac12 \kappa_{\Xi^-}\right),
\label{last}
\end{eqnarray}
where $\kappa_B$ is the anomalous magnetic moment corresponding 
to the baryon $B$.
These relations have corrections of order ${\cal O} (m^2_{s})$
and ${\cal O} (m_s/N_c)$ which are assumed to be small.
 In exact SU(3) symmetry these relations reduce again to the SU(3)
relations shown in Table I.  
 In Table~II we compare the results with SU(3) symmetry breaking
to those in SU(3) symmetry.  The experimental data for $f_2(0)$
are taken from Ref.\cite{Dworkinetal}
for $\Lambda\rightarrow p  e^- \bar{\nu}$
and from Ref.\cite{Hsuehetal}
for $\Sigma^- \rightarrow ne^- \bar{\nu}$.
Incorporating experimental values for magnetic moments taken 
from \cite{PDG96} into formulae in Table I and those in Eq.(\ref{last}),
respectively, we obtain the $f_2(0)/f_1 (0)$ ratios for seven different
channels~\footnote{Note that though we use the SU(3) symmetric expressions
to obtain the first column in Table II the results nevertheless include 
a part of the SU(3) symmetry breaking through the experimental data.
 However, by doing that we can see at least the effect of SU(3) symmetry
breaking within the framework of the $\chi$QSM.}.
Let us first compare the first two columns.  Apart from the
$|\Delta S|=0$ modes for which the effect of SU(3) symmetry breaking
is observed in around $10\ \%$, we can see the comparably large 
effect of the SU(3) symmetry breaking.  Considering the 
SU(3) symmetry breaking inherent already in experimental magnetic
moments, one can say that the effect of SU(3) symmetry breaking is
even larger.  From the comparison of the third and fourth columns,
we can find the effects of SU(3) symmetry breaking.
The effects on $f_2(0)/f_1(0)$ are noticeably large in almost
every decay mode.  In particular, the deviation from SU(3) symmetry
appearing in the $\Sigma^- \rightarrow n e^- \bar{\nu}$ mode is remarkable.
 Indeed, the effects of SU(3) symmetry breaking pull
$f_2(0)/f_1(0) (\Sigma^-\rightarrow n)$ 
drastically down from its SU(3) symmetric value, 
so that it turns out to be in good agreement with the data~\cite{Hsuehetal}. 

The expressions of $g_1(0)$ for the case of SU(3) symmetry breaking
can be obtained similarly to the SU(3) symmetric case (see Table I).  
For convenience, we define the baryonic axial constants as follows:
\begin{equation}
g^B_{A} \;=\; \left\langle B\left| \left(g^{(3)}_A + \frac{1}{\sqrt{3}}
g^{(8)}_A \right)\right|B\right\rangle.
\label{Eq:axial}
\end{equation}
Then we can write the expressions for $g_1(0) (B_1\rightarrow B_2)$
similar to Eq.(\ref{last}):
\begin{eqnarray}
g_1^{(n \rightarrow p)}(0) &=& \frac12\left(g^{p}_A-g^{n}_A\right),
\nonumber\\
g_1^{(\Sigma^- \rightarrow \Sigma^0)}(0) &=& 
\frac{1}{2\sqrt{2}}\left(g^{\Sigma^+}_A-g^{\Sigma^-}_A\right), 
\nonumber\\
g_1^{(\Sigma^\pm \rightarrow \Lambda)}(0)&=& \mp \frac12\sqrt{\frac23}
\left(-g^n_{A} + \frac12 g^{\Sigma^0}_A 
- g^{\Xi^0}_A + \frac32 g^{\Lambda}_A\right), \nonumber \\
g_1^{(\Lambda \rightarrow p)}(0)&=& \frac{1}{2\sqrt{6}}
\left(g^n_{A}+\frac52 g^p_{A}+ \frac12 g^{\Sigma^-}_A
-3g^{\Lambda}_A -g^{\Xi^-}_A\right),\nonumber\\
g_1^{(\Sigma^- \rightarrow n)}(0)&=& \frac12
\left(g^n_{A}+\frac12 g^p_{A} - g^{\Sigma^-}_{A}
-\frac12 g^{\Sigma^+}_A\right), \nonumber \\
g_1^{(\Xi^- \rightarrow \Lambda)}(0) &=& \frac{1}{2\sqrt{6}}
\left( g^p_{A}-\frac12 g^{\Sigma^+}_A+3g^{\Lambda}_A -
g^{\Xi^0}_A -\frac52 g^{\Xi^-}_A\right),\nonumber\\
g_1^{(\Xi^- \rightarrow \Sigma^0)}(0)&=& \frac{1}{2\sqrt{2}}
\left( g^{\Sigma^+}_A +\frac12 g^{\Sigma^-}_A -
g^{\Xi^0}_A - \frac12 g^{\Xi^-}_A\right),
\end{eqnarray}
Making use of Eq.(\ref{Eq:g1}), we obtain the sum rule 
between six different decay modes 
\begin{equation}
g_1^{(n \rightarrow p)}=
\frac12\left(\sqrt2 g_1^{(\Sigma^- \rightarrow \Sigma^0)}
+\sqrt6 g_1^{(\Sigma^+ \rightarrow \Lambda)}-
\sqrt6 g_1^{(\Lambda \rightarrow p)}+
 g_1^{(\Sigma^- \rightarrow n)}-
 2\sqrt2  g_1^{(\Xi^- \rightarrow \Sigma^0)}\right).
 \label{ga}
\end{equation}
In order to verify Eq.(\ref{ga}), more accurate experimental 
data are required than presently available.
 In the SU(3) limit the right hand side of
Eq.(\ref{ga}) becomes $D+F$ in accordance with the Cabibbo theory 
\cite{Cabibbo}.

In Table III, the results of $g_1(0)/f_1(0)$ with SU(3) symmetry breaking 
are compared to those in SU(3) symmetry.  The effect of SU(3) symmetry 
breaking is measured in $5\ \sim \ 10\ \%$, which is not that strong.
Compared to the case of $f_2(0)/f_1(0)$, the effect of SU(3) symmetry
breaking is rather soft.  This is partly due to the fact that
in the axial channel the mass differences do not come into play,
which is somewhat in line with the argument of Ref.~\cite{Ratcliffe}.  
The results predicted by the $\chi$QSM are in good agreement with
the experimenta data~\cite{PDG96} within about $15\%$ which is a typical
predictive power of the model.  In particular, the results 
agree with the data remarkably in $|\Delta S| = 1$ channels. 

It is also interesting to see that 
in the present model the ratio of $g_1/f_1$ between 
$\Lambda\rightarrow pe^-\bar{\nu}$ decay and
$\Sigma^-\rightarrow ne^-\bar{\nu}$ decay is well reproduced
\begin{equation}
\frac{g_1/f_1 (\Lambda\rightarrow pe^-\bar{\nu})}
{g_1/f_1 (\Sigma^-\rightarrow ne^-\bar{\nu})}\;=\;-2.28\;
({\rm Exp:}\; -2.11\pm 0.15)
\end{equation}
It was pointed out that this ratio is {\em a priori} constrained
to $-3$ in quark models with SU(6) symmetry~\cite{Lipkin92,Schlumpf},
which is noticeably larger than the experimental value.

It is also interesting to compare the present results with those from
the Skyrme model with vector mesons~\cite{ParkWeigel}.
 Ref.~\cite{ParkWeigel} presented the $g_1/f_1$ ratio in five different
channels.  Except for the $\Lambda\rightarrow p$ mode, 
the present model seems to be far better than Ref.~\cite{ParkWeigel}.
For example, we obtain the $|g_1/f_1|=0.31$ for the 
$\Sigma^- \rightarrow n$ mode comparable to the
experimental data $0.34\pm0.017$, while Ref.~\cite{ParkWeigel} 
yields $0.24$.    

From Eq.(\ref{g2}), we can obtain the ratio $g_2(0)/f_1(0)$ 
in terms of $\beta_1$ and $\beta_2$.   In Table IV, the expressions 
and numerical results for them are listed.
Numerically, $\beta_2$ is much larger than $\beta_1$, 
which explains why the $g_2/f_1$ for the
$\Xi^-\rightarrow \Sigma^0$ mode turns out to be 
much greater than those for the other modes.    
Our results are compared with results of an SU(6) relativistic quark 
model~\cite{Kellett} as well as with those of a light-front 
relativistic quark model~\cite{Schlumpf}.  We see that our results
are close -- within factor of 2 -- to the results of Ref.~\cite{Schlumpf},
whereas differ by almost order of magnitude from those of 
Ref.~\cite{Kellett}.  
 
\section{Summary and conclusion}
The aim of the present work has been to investigate baryon semileptonic
decays within the framework of the chiral quark-soliton model ($\chi$QSM).
In particular, the role of SU(3) symmetry breaking in baryon 
semileptonic decays have been discussed.  
Based on the recent studies on the octet magnetic moments and 
axial vector constants, we have obtained the ratios $f_2/f_1$ and
$g_1/f_1$ without and with SU(3) symmetry breaking, respectively.
In exact SU(3) symmetry, we have shown that $g_2$ vanishes in
the present model, as it should be.  We have discussed also that
for the $F/D$ ratio the $\chi$QSM ($0.61$) interpolates the 
Skyrme model ($5/9$) and the NRQM ($2/3$).  

It was found that the effect of SU(3) symmetry breaking 
differs in different channels and modes.     
In general, SU(3) symmetry breaking contributes strongly to
the ratio of the vector coupling constants $f_2(0)/f_1(0)$ while
it does not much to the ratio $g_1(0)/f_1(0)$.  
In addition, we have evaluated the ratio $g_2/f_1$.  This is
the first calculation of the $g_2/f_1$ in soliton models.  
 The results come out to be small
except for the $\Xi^-\rightarrow \Sigma^0$ mode.
Due to lack of experimental data, 
the values of $g_2(0)$ we calculated are predictions. 

\section*{Acknowledgment}
This work has partly been supported by the BMBF, the DFG
and the COSY--Project (J\" ulich).  M.V.P. and M.P. have been 
supported by Alexander von Humboldt Stiftung.  

\begin{table}
\caption{The expressions of $f_1(0)$, $f_2(0)$ and $g_1(0)$
in exact SU(3) symmetry.
The $\kappa_p$ and the $\kappa_n$ denote the anomalous magnetic moments 
of the proton and the neutron, respectively.  } 
\begin{tabular}{cccc}
Decay mode & $f_1(0)$ & $f_2 (0)$&$g_1(0)$  \\ \hline
$n\rightarrow p$ & $1$ & $\frac12 (\kappa_p - \kappa_n)$&$F+D$  \\
$\Sigma^-\rightarrow \Sigma^0$ & $\sqrt{2}$ & $\frac{1}{\sqrt{2}}
(\kappa_p + \frac12 \kappa_n)$&$\sqrt{2}F$ \\
$\Sigma^{\pm} \rightarrow \Lambda$ & $0$ & 
$\pm \sqrt{\frac38} \kappa_n$ &$\mp\sqrt{\frac23}D$ \\
$\Lambda\rightarrow p$ & $\sqrt{\frac32}$ & 
$\frac12\sqrt{\frac32}\kappa_p$ &$\sqrt{\frac32}(F+D/3)$ \\
$\Sigma^-\rightarrow n$ & $1$ & $\frac12 (\kappa_p + 2\kappa_n)$&$F-D$ \\ 
$\Xi^-\rightarrow\Lambda$ & $\sqrt{\frac32}$ & 
$\frac12\sqrt{\frac32}(\kappa_p + \kappa_n)$ &$\sqrt{\frac32}(F-D/3)$\\
$\Xi^-\rightarrow \Sigma^0$ & $\frac{1}{\sqrt{2}}$ &
$\frac{1}{2\sqrt{2}} (\kappa_p - \kappa_n)$&$\sqrt{\frac12}(F+D)$
\end{tabular}
\end{table}
\begin{table}
\caption{The results of $f_2(0)/f_1(0)$. The column under 
${\rm SU(3)}_{\rm sym}$ lists the case of exact SU(3) symmetry with
the experimental data of $\kappa_p$ and $\kappa_n$ while 
the next column shows the case of broken SU(3) symmetry by $m_s$ with
the experimental data of $\kappa_B$~\protect{\cite{PDG96}}.  
 The third column presents
the numerical results of $f_2(0)$ in the $\chi$QSM with the constituent
quark mass $M=420$ MeV, while the fourth one shows those of $f_2(0)$
in the $\chi$QSM with the same constituent quark mass.
 The experimental data are taken from Ref.{\protect\cite{Dworkinetal}} 
for $\Lambda\rightarrow p  e^- \bar{\nu}$ and from 
Ref.\protect{\cite{Hsuehetal}}
for $\Sigma^- \rightarrow ne^- \bar{\nu}$.}
\begin{tabular}{cccccc}
 decay mode  & ${\rm SU(3)}_{\rm sym}$ 
&  ${\rm SU(3)}_{\rm br}$ &  $\chi{\rm QSM}_{\rm sym}$ &
$\chi{\rm QSM}_{\rm br}$ & Exp  \\
\hline
$n \rightarrow p$& $1.853 $ & $1.853$ & $1.41$ & $1.58$ &$--$ \\
$\Sigma^- \rightarrow \Sigma^0$ & $0.418 $ & $0.516 \pm 0.012$ 
& $0.25$ & $0.43$ &$--$ \\
$\Sigma^{-} \rightarrow \Lambda$ & $1.435$\tablenotemark[1]  
& $1.625 \pm 0.011$\tablenotemark[1] 
&$0.95$\tablenotemark[1]  & $1.33$\tablenotemark[1] &$--$  \\
$\Lambda \rightarrow p$ & $0.896 $ & $0.787\pm 0.004 $
&$0.64$ & $0.74$ & $0.15\pm 0.30$ \\
$\Sigma^- \rightarrow n$ & $-1.017 $ & $-1.010 \pm 0.016$
&$-0.92$ & $-1.18$& $-0.96\pm 0.07\pm 0.13$ \\
$\Xi^- \rightarrow \Lambda$ & $-0.060 $ & $-0.093 \pm 0.006$
&$-0.14$ & $-0.18$ & $--$\\
$\Xi^- \rightarrow \Sigma^0$ & $1.853 $ & $1.725 \pm 0.011$
&$1.41$ &$2.06$&$--$\\
\end{tabular}
\tablenotetext[1]{Instead of $f_2/f_1$, we list $\sqrt{\frac32}f_2$.}
\end{table}
\begin{table}
\caption{The results of $g_1(0)/f_1(0)$. The column under 
$\chi {\rm QSM}_{\rm sym}$ lists the numerical results in 
exact SU(3) symmetry while in the next column shows the case of 
broken SU(3) symmetry by $m_s$.
The constituent quark mass $M=420$ MeV is used.
Most experimental data are taken from Particle Data Group
\protect{\cite{PDG96}}.  The data for the $\Sigma^{-} \rightarrow \Lambda$
mode are taken from \protect{\cite{BGHORS1}} while that for the
$\Xi^- \rightarrow \Sigma^0$ mode are from \protect{\cite{BGHORS2}}.}
\begin{tabular}{cccc}
 decay mode  &  $\chi{\rm QSM}_{\rm sym}$&$\chi{\rm QSM}_{\rm br}$
& Exp \\
\hline
$n \rightarrow p$& $1.33$ & $1.42$ & $1.2573\pm 0.0028$ \\
$\Sigma^- \rightarrow \Sigma^0$ & $0.50$ & $0.55 $ & $--$ \\
$\Sigma^{-} \rightarrow \Lambda$
& $0.83$\tablenotemark[1]  & $0.91$\tablenotemark[1]  
&$ 0.720\pm 0.020$\tablenotemark[1]  \\
$\Lambda \rightarrow p $ & $0.78$ & $0.73$ & 
$0.718 \pm 0.015$ \\
$\Sigma^- \rightarrow n$ & $-0.33$ & $-0.31$ & 
$-0.340 \pm 0.017$\\
$\Xi^- \rightarrow \Lambda$ & $0.23$ & $0.22$ & 
$0.25\pm 0.05$\\
$\Xi^- \rightarrow \Sigma^0$ & $1.33$ & $1.29$ & 
$1.25^{+ 0.14}_{-0.16}$\\
\end{tabular}
\tablenotetext[1]{Instead of $g_1/f_1$, we list $\sqrt{\frac32}g_1$.}
\end{table}
\begin{table}
\caption{The induced pseudotensor coupling constants ratio $g_2(0)/f_1(0)$.
The results are compared with Refs.~\protect{\cite{Schlumpf,Kellett}}.}
\begin{tabular}{ccccc}
 decay mode  & expression  & $\chi$QSM & Schlumpf & Kellett\\
\hline
$n \rightarrow p$&$0$& $0$ & $0$ & $0.29$   \\
$\Sigma^- \rightarrow \Sigma^0$ & $0$ & $0$ & $0 $ &$--$ \\
$\Sigma^{-} \rightarrow \Lambda$ & 
$  -\frac{4m_s}{5\sqrt{3}}  M_{\Sigma^-} \beta_2$ \tablenotemark[1]  
& $-0.029$\tablenotemark[1] & $0$\tablenotemark[1] 
& $0.18$\tablenotemark[1]  \\
$\Lambda \rightarrow p $ &$\frac{8m_s}{5\sqrt{3}} M_{\Lambda}
\left(\beta_1 + \frac12 \beta_2\right)$ & $0.046$& $0.023$&$0.25$  \\
$\Sigma^- \rightarrow n$ & 
$-\frac{4m_s}{15\sqrt{3}}M_{\Sigma^-}\left(3\beta_1 + \beta_2\right)$
 & $-0.020$& $-0.007$& $-0.09$  \\
$\Xi^- \rightarrow \Lambda$ & $\frac{2m_s}{5\sqrt{3}}M_{\Xi^-}\beta_1$
 & $0.006$& $0.008$ & $--$ \\
$\Xi^- \rightarrow \Sigma^0$ & $\frac{2m_s}{15\sqrt{3}}M_{\Xi^-} 
\left(21 \beta_1 + 16 \beta_2\right)$  & $0.125$ & $0.04$ & $--$\\
\end{tabular}
\tablenotetext[1]{Instead of $g_2/f_1$, we list $\sqrt{\frac32}g_2$.}
\end{table}

\end{document}